\documentclass{emulateapj}
\usepackage{graphicx}

\shorttitle{EXPECTED BLACK HOLE SPINS IN BINARIES}
\shortauthors{O'Shaughnessy et al.}

\begin{document}
\title{Bounds on Expected Black Hole Spins in Inspiraling Binaries}
\author{ R.\ O'Shaughnessy$^{1}$, J. Kaplan$^{1}$, V.\ Kalogera$^{1}$,  \& K.\ Belczynski$^{2}$}

\altaffiltext{1}{Northwestern University, Department of Physics and
  Astronomy, 2145 Sheridan Road, Evanston, IL 60208, USA}
\altaffiltext{2}{Tombaugh Fellow, Department of Astronomy, New Mexico State University, Las Cruces, NM 88003}
\email{oshaughn, j-kaplan-1, vicky@northwestern.edu and kbelczyn@nmsu.edu}

\begin{abstract}
As a first step towards understanding the angular momentum
evolution history of black holes in merging black-hole/neutron-star
binaries, we perform population synthesis calculations to track the distribution of accretion histories of compact objects in such binaries.  
We find that there are three distinct processes which can possibly
contribute to the black-hole spin magnitude:
a birth spin for the black hole, imparted at either (i) the collapse of a massive progenitor star to a black hole or
(ii) the accretion-induced collapse of a neutron star to a black hole;
and (iii) an accretion spin-up when the already formed black hole [via
(i) or (ii)] goes
through an accretion episode (through an accretion disk or a
common-envelope phase).  Our results show that, with regard to
accretion-induced spinup in merging BH-NS binaries [method (iii) above], only 
 {\em accretion episodes associated with
common-envelope phases and hypercritical accretion rates} occur in the formation history of merging black hole/neutron star binaries. 
Lacking unambiguous experimental information about BH birth spins
[i.e., regarding the results of processes (i) and (ii)], we choose two fiducial
values for the BH birth angular momentum parameter $a=J/M^2$,
consistent with observations of (i) NS birth spins ($a\simeq0$) and (ii) X-ray binaries ($a=0.5$).  Using these two fiducial values and  
a conservative upper bound on the specific
angular momentum of accreted matter, 
we discuss the expected range of black hole spins in the binaries of interest.
We conclude with comments on the significance of these results for ground-based gravitational-wave searches of inspiral signals from black hole binaries.

 \end{abstract}

\keywords{binaries:close ---  stars:evolution ---  stars:neutron --
  black hole physics -- gravitational waves -- relativity -- X-rays:binaries}

\maketitle

\section{Introduction}
Gravitational-wave astronomy is a rapidly-developing field, with many
ground-based interferometric detectors (LIGO, VIRGO, GEO, TAMA) in
place and taking data.
These detectors are sensitive to (among other things) the waves
emitted from the late stages of inspiral and merger of compact
binaries with neutron stars (NS) and black holes (BH). 
Under certain conditions  \citep[as discussed in][and references therein]{ACST,Mimic,Spiky3,BCVspin2}  
the BH spin in high-mass ratio binaries such as BH-NS binaries can significantly modulate the
emitted inspiral waveform. 
While the more complex resulting waveforms can pose challenges for 
detection strategies, they also enable us to 
empirically determine the spin of rapidly-rotating black
holes, if they exist.
We undertake the present study of the expected BH spin distribution for two reasons: (i) to examine whether special search methods for spinning compact objects for optimal detection efficiency are necessary from an astrophysical point of view; (ii) to provide the first steps for a theoretical understanding of the origin and magnitude of  BH spins in BH-NS  binaries  that will be
useful when such BH spins are empirically constrained in the future.
 
In Sec.~\ref{sec:ps}, we briefly describe the population synthesis
calculations we use to explore a broad range of possible scenarios that lead to the formation of BH-NS binaries that can merge within a Hubble time (hereafter referred to as ``merging'').  We summarize and explain the common formation channels for these binaries and 
describe the mass accretion history associated with them.
In Sec.~\ref{sec:birth}, we review the observational evidence for BH
birth spin drawn from observations of isolated NS and of X-ray
binaries.  Based on this evidence, we employ two fiducial choices for
BH birth spin parameter $a=J/M^2$, i.e., $a=0$ (nonspinning) or
$a=0.5$ (moderate spin).
Given an understanding of the mass accretion history and a choice for the initial
BH spin , we could in principle determine the final 
black hole spin if we knew the specific angular momentum of the
material accreted by BH in these systems.   This factor remains substantially uncertain.
Thus, in Sec.~\ref{sec:spinup} we use a \emph{conservative}
model for the specific angular momentum, combined with the
distribution of accretion histories and birth spins, 
to 
limit the rate at which high-spin black holes occur in merging BH-NS binaries.

\section{Accretion History of
   Black Holes in BH-NS binaries}
   \label{sec:ps}

To generate and evolve stellar populations until double compact-object
formation occurs, we use the \emph{StarTrack} code first developed by
Belczynski, Kalogera, and Bulik
(2002) [hereafter BKB] and recently significantly updated and tested as described in detail in \citet{StarTrackUpdates}.
As with all binary population synthesis codes, 
\emph{StarTrack} encapsulates the
residual physical uncertainty in evolutionary processes of stars in
a number of  parameters, several of which can
significantly affect the statistics of BH-NS mergers.  In order to have a sample of merging BH-NS drawn from a large archive of
astrophysically \emph{a priori} plausible models, we generated an archive of
accumulated results from a broad swath of 
plausible models for population synthesis.   
As described in 
O'Shaughnessy, Kalogera, and Belczynski (2004) [hereafter denoted OKB], 
the archive was generated using flat probability distributions for seven of the most significant parameters:
three parameters describing the supernova
kick magnitude distribution (assumed to consist of two maxwellian distributions
with dispersions $\sigma_1\in[0,200]$km/s and
$\sigma_2\in[200,1000]$km/s and relative weight of small to large kicks $s\in[0,1]$); the effective common-envelope efficiency $\alpha \lambda \in[0,1]$; the 
stellar wind strength $w\in[0,1]$; the power-law index controlling the distribution of the mass
of the companion relative to the mass of the primary $r\in[-3,0]$; and the fraction of mass accreted (as opposed to lost from the binary) during non-conservative mass
transfer episodes $f_a\in[0,1]$.
Unfortunately, as discussed in more detail in \citet{PSconstraints},
we cannot  succinctly describe the effect of these seven
parameters on the formation channels and rates of BH-NS systems;
physical predictions depend in a \emph{highly correlated way on all
seven parameters}.
As in OKB, we fixed the remaining population synthesis model
parameters  to physically reasonable values, such as setting  the
maximum  neutron star mass to $M_{NS,max}=2 M_\odot$ and the
metallicity  to solar metallicity $Z=0.02$ (see parameters of model A in~BKB for more details).
From this archive we identified each merging
BH-NS binary along
with the details of its evolutionary history.   Our results are
summarized in Table~\ref{tab:main}.

Compact binaries tight enough to merge through the emission
of gravitational waves within the Hubble time have
progenitors which usually interacted strongly (e.g. through mass transfer)
earlier in their evolution.  For example, a common mode of interaction
is conventional mass transfer through Roche-lobe overflow and disk accretion. 
However, for merging BH-NS binaries, we find in our simulations that
almost all of them have experienced a more dramatic phase  of dynamically
unstable mass transfer and common envelope (CE) evolution (see
Table~\ref{tab:main}).  During such phases it is expected that {\em
hypercritical} accretion becomes possible.  In this form of accretion
(HCE), discussed in more detail in \citet{BrownHCE1},
\citet{BrownHCE2}, and \citet{FryerBenzHerant}, a compact object
spirals in through the envelope of its companion, rapidly accreting
matter at highly super-Eddington (for photons), neutrino-cooled rates.
As a result of this process, most of the companion's envelope is
ejected, bringing the post-CE binary very close together.
While the detailed accretion  onto the compact object is complex and
ill-understood, since we know the whole envelope will be lost
during this phase, a straightforward application of the conservation of energy and mass during the quasicircular spiral-in process can 
determine the final orbit, the final binary masses, and the mass of
the ejected material; see Appendix A of \citet{StarTrack} for
details. Nevertheless we stress that our treatment of HCE is
essentially a simple, semi-analytical model, which is by no means
guaranteed to accurately describe the quantitative effect of the HCE
phase.

\begin{deluxetable}{crr|rr}
\tablecolumns{6}
\tablecaption{Statistics for evolutionary channels}
\tablehead{ 
&\colhead{Channel} & & \colhead{n}  & \colhead{AIC}} \startdata
{\rm (MT+)SNa}  & HCE(b$\rightarrow$a) & {\rm (MT+)SNb}&
$4310$ & $3656$ 
\\ 
{\rm (MT+)SNb}  & HCE(a$\rightarrow$b) & {\rm (MT+)SNa}& 
$543$ & $543$ 
\\ 
{\rm (MT+)SNx}  &  & {\rm MT+SNy}& 
$319$ & $0$ 
\\ 
{\rm (MT+)SNx}  &  & {\rm SNy}& 
$263$ & $0$ 
\\ \hline  
{\rm total:}      & & & 
5435 &
\enddata
\label{tab:main}
\tablecomments{This table summarizes all evolutionary channels followed by systems which formed merging BH-NS binaries in our simulations. The first three columns describe the evolutionary channel (where SNx indicates a supernovae
of either $a$ the primary or $b$ the secondary, HCE indicates a
hypercritical common envelope phase, and MT indicates a stable
Roche-lobe-overflow mass transfer; parentheses indicate an optional
feature of the channel); the fourth column provides the
number of merging BH-NS binaries which passed through this channel;
and the fifth lists the number which  undergo
accretion-induced-collapse (i.e., the BH forms during the
common-envelope phase from an accreting neutron star).   Most
systems formed via a hypercritical common envelope (HCE) phase onto a
compact object; this compact object is occasionally already BH, but
usually is a  NS which 
undergoes accretion-induced collapse (AIC) to a BH.  A smaller number
of systems (582 of 5435) do not undergo HCE at all and interact either
weakly (through winds) or not at all between the two supernovae.  
} 
\end{deluxetable}

\begin{figure}
\includegraphics{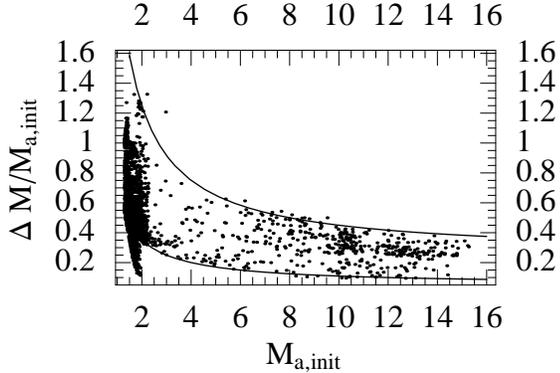}
\caption{\label{fig:dM}
For systems that undergo HCE evolution
(see  Table~\ref{tab:main}), this plot shows 
  $(M_{\text{final}}-M_{\text{inital}})/M_{\text{initial}}$
 versus $M_{\text{initial}}$, where $M_{\text{initial}}$ and 
 $M_{\text{final}}$ are the mass of the BH progenitor just after its
 supernovae and at the end of its evolution, respectively.
 In other words, they show the relative increase in mass of the
  compact object $\Delta M_{\text{HCE}} {}/M$ due to hypercritical common
  envelope accretion.  The upper solid line shows the empirical upper bound
  given in Eq.~(\ref{eq:HorribleGuess}); the two vertical lines
  shows our assumed lower and upper limits on NS mass, $M=1.3
  M_\odot$, $2 M_\odot$. 
 }
\end{figure}

More specifically, as described in Table~\ref{tab:main}, the vast
majority (all but $582$, out of $5435$) of BH-NS binaries seen in our simulations form through an evolutionary channel
that involves an HCE phase.
In these channels,
 after an optional mass transfer phase, one star explodes,
the compact remnant spirals through and strips the envelope of its
companion, and then the companion explodes.
Black holes form either
immediately after the first core-collapse event or through the
accretion-induced collapse of a NS that forms in the first supernova
event. Regardless of the nature of the first compact object (BH or NS),
it most typically experiences HCE evolution, and accretes a large
amount of matter (potentially equal to its birth mass) during the
brief HCE phase.  Figure~\ref{fig:dM} shows the fraction of the
initial compact object's mass $M_{\rm init}$ that is accreted during the HCE phase.  
Based on the sample we have, we
estimate an upper bound on the fractional mass accreted as a function
of the initial mass to be
\begin{equation}
\label{eq:HorribleGuess}
\left(\Delta M/M \right){}_{max} {} \approx 0.25+2 M_\odot/M_{\rm init} \; .
\end{equation}
This bound is valid for all channels.\footnote{We expect that systems which transfer more mass than this limit are brought so close that they merge during their common-envelope phase.}
For systems which form their BH immediately after the first supernova
    (i.e., no accretion-induced 
collapse, so $M_{\rm a,init}>2 M_\odot$) we find an empirical \emph{lower}
bound of
\begin{equation}
\label{eq:HorribleGuess2}
\left(\Delta M/M \right){}_{max} {} \approx 0.05+0.6 M_\odot/M_{\rm init} \; .
\end{equation}
This latter bound may be understood as  a requirement that the BH-NS system
merge within the simulation time.  Given that the second  supernova
will rarely  
further tighten the orbit, if the binary is to merge within
10 Gyr,   then the orbit must be smaller  than roughly $3.4 R_\odot
\left[(M_1/M_\odot)(M_2/M_\odot)(M_1+M_2)/M_\odot\right]^{1/4}$ after
the common envelope phase.  Using the common-envelope evolution
equations presented in Appendix A of \citet{StarTrack} (with any
choice for $\alpha_{\rm CE}\times\lambda$), we find this requirement
approximately translates into the lower bound presented above.

Merging BH-NS can form through other channels, as shown in
Table~\ref{tab:main}.  However, these channels involve substantially
less mass transfer onto the BH: for the HCE-related channels,
the \emph{smallest} mass increase was $0.22
M_\odot$, to be contrasted with the \emph{largest} disk-mediated mass
transfer,
$0.007 M_\odot$ (i.e., from the channel SNx+MT+SNy).  As described in Section~\ref{sec:spinup}, this
incredibly small disk-mediated mass transfer cannot significantly
modify the birth spin of the BH.   Therefore, we neglect these
channels in subsequent discussion. 

\emph{Caveat: Branching ratios}: Unfortunately, our branching ratios for various evolutionary channels are not significant. When exploring different population synthesis parameter combinations, we  normalized so each population synthesis run contributes a roughly fixed number of merging NS-NS binaries to the archive.  As a result, models associated with low NS-NS merger rates but high BH-NS merger rates should contribute disproportionately more merging BH-NS to our archive, and vice-versa.   Moreover, even if each model contributed equally to the total number of BH-NS systems, our net numbers would then reflect the \emph{mean branching ratios} over all population synthesis models considered, rather than the branching ratio of, e.g., the most common model.   Finally, as discussed in \citet{PSconstraints}, not all population synthesis models we considered are equally compatible with observations.  Nonetheless given the current statistics, we fully expect that most BH-NS systems will form via a hypercritical common envelope phase.

\section{Birth spins of BH in BH-NS binaries}
\label{sec:birth}
No direct measurements of BH birth spins (or even BH spins) exist yet.
A priori the BH birth spin cannot be constrained
beyond the most fundamental level (i.e., to avoid naked singularities,
we must have low angular momentum $J$: specifically, 
$a\equiv J/M^2\le 1$).
 For example, the collapse of a
slightly-hypermassive NS rotating at breakup will, on dimensional
grounds alone, produce a BH of spin of order half of maximal [$a\sim
0.5$] should it collapse; more realistic computations which allow for potential
differential rotation support could push this value even higher
\citep[see, e.g.,][]{Duez1,ShapiroPolytropicCollapse,Hawke1}.

Lacking unambiguous theoretical guidance for BH birth spins in merging
BH-NS systems, and lacking direct observations of BH spin,  we employ
evidence for birth spin in similar systems to guide our choices for BH
birth spin.
On the one hand,  on the basis of our understanding of NS birth spins
and spinup in NS-NS binaries, we expect \emph{low} BH birth spins ($a=0$).
However, on the basis of suggestive observations of X-ray binaries, we
suspect \emph{moderate} BH birth spins could occur ($a=0.5$)

\subsection{BH birth spin estimates from NS observations}
The same processes that produce BH birth spin (i.e. core
collapse and accretion-induced spinup) also determine the spins of the
well-observed NS population.  Observations of the NS population can
thus potentially provide us with some constraints on BH birth spins.

\emph{BH birth spin from core collapse}: We expect we can estimate the
birth spins $a=J/M^2$ of BHs, and particularly low-mass BHs, through the
estimated birth angular momenta $J$ of young NS.  For BHs and NSs of
comparable mass, the collapse process should be nearly identical;
therefore, the collapse product should have very similar $J$ and $M$.
Assuming no other process intervenes [for example, as discussed in
\citet{rmode1}, r-mode damping is not expected to significantly change
the NS angular momentum of young NS], NS will spin down
   electromagnetically, and thus NS spins at birth can be estimated
   from the observed pulsar NS sample.
In the past few years a number of
studies of observed radio pulsars have estimated the NS spin periods
at birth \citep[see][and references therein]{Lorimer1,Kramer1,Mig1} in the range of $10-140$\,ms. These
values are significantly slower than break-up spins (for typical NS
equations of state) and correspond to $a\simeq 0.005-0.02$ (assuming
a NS radius of $10$\,km and rigid rotation).

\emph{BH birth spin from accretion-induced collapse}: Our simulations
show that a significant fraction (i.e., 4199 out of 5435 binaries) of the BH in merging BH-NS systems
were originally NS that experienced HCE and collapsed into
BH. Therefore birth spins for this BH class are related to the spin-up
of NS during a CE phase. Our current understanding of this spin-up
process is quite limited. 
However, we can
obtain guidance from the mildly recycled pulsars in known double
neutron star binaries. These pulsars are believed to have been spun up
during a CE phase (where collapse to a BH was avoided though) and it
is evident that they are not spinning close to break-up: the fastest
known pulsar in a double NS binary is PSR J0737-3039A spinning at
$\simeq 20$\,ms or else having $a=0.01$ \citep{0737}. 
Therefore post-CE NS spins appear to correspond to $a$ no larger than
$a=0.01$.

Based on the observational considerations discussed above, we would
expect the birth BH spin to be negligible (i.e., $a\lesssim 0.03$).
However, a number of uncertainties in our modeling suggest that very
different values of $a$ could be equally plausible.

\emph{Uncertainties in collapse model}:  For black holes that form
from the supernova of a massive star, 
 we assume that the birth spin of these BHs are comparable to
the birth spin of comparable-mass NSs, and we then assume that those
NS birth spins may be estimated from electromagnetic (pulsar) spindown
from present-day pulsars.   
However, most of these low-mass black holes form through fallback of
post-supernova ejecta onto the hole; if the post-supernova fallback material
carries significant angular momentum and can spin up the nascent BH to
high values of $a$, even if the protoneutron star is not spinning with
a period shorter than $\simeq 10$\,ms.  Further, if young protoneutron
stars can spin down rapidly through other mechanisms, such as r-modes
\citep{rmode1}, straightforward electromagnetic-based extrapolations
may significantly underestimate the protoneutron star spin.  If either
of these mechanisms took place, then the birth BH spin could be
significantly larger than the naive estimate we outlined above.

\emph{Uncertainties in HCE model}: The known pulsars in double NS binaries have avoided
accretion-induced collapse into BH. Those that do not avoid this
collapse must accrete a higher amount of mass and therefore it is
still possible that they get spun-up to shorter spin periods and
therefore lead to BH formed with higher $a$ values than expected.

Given these uncertainties, we cannot be sure whether the relevant BH
birth spins are negligible (i.e., $a=0$) or large, comparable to the
break-up spin of a NS (i.e., $a\sim 0.5$).

\subsection{Black hole Spins  in X-Ray Binaries}
Black hole X-ray binaries (XRBs) offer
the possibility of \emph{directly} measuring the spin of a compact object, since they involve a highly relativistic accretion flow in
the strong field of a BH. 
Two techniques are prominent in the literature: interpreting quasiperiodic
oscillations and fitting iron line profiles.

Quasiperiodic oscillations are believed to be modes of the inner,
highly relativistic regions of the BH accretion disk.   As such, their
frequency is expected to be related to the Kepler frequency of the
inner edge of the disk, which in turn is intimately connected with the
properties of the black hole.  Measurements of high-frequency QPOs
that are
inconsistent with the innermost stable circular orbit of a nonspinning
black hole, as by \citet{Strohmayer} and \citet{Remillard}, have been interpreted as
evidence for BH spin.   
Moreover, evidence suggests the spin suggested 
by these measurements is \emph{not} merely a relic of past accretion.  
Unfortunately, QPO frequencies cannot be unambiguously translated to
black hole angular momenta; their interpretation depends strongly on
disk mode modelling \citep[see, e.g.][]{Abramowicz,Rezolla}, with
results that vary from $a\approx 0.1$ to $a\approx 1$ depending on
the  assumptions used.
However, even if the QPO
interpretation is correct,  the associated BH spin
estimates need not reflect BH birth spins; such spins are most probably reached via  long-term disk accretion, a process that is not relevant to BHs in BH-NS merging binaries. 

Iron line profiles potentially offer in principle a more direct 
probe into the inner disk rotation profile and, thus, the BH
spin.  The iron line can show strong Doppler shifts (due to
rotation) and redshifts (due to strong gravity); in particular, very
rapidly spinning BHs should show asymmetric red ``tails''.
While pioneered for use with AGN, this same technique has been applied
to BH XRBs \citep[see, e.g.,][]{Miller1,Miller2,Miller3}.
We treat the results,  several claims of \emph{extreme} spins $a>0.8$,
with great caution however.  On the one hand, these systems are strongly model
dependent, and many physically relevant details  \citep[e.g., proper
accounting of light bending near the hole; see][]{Beckwith} have yet
to be included in the models.  On the other hand, just like in the case of QPO interpretations, the estimated spins may be entirely due to the disk accretion in the XRB and therefore unrelated to the birth BH spin.

\subsection{Birth Black Hole Spins Adopted }
To summarize, based on observations of isolated NS we believe the
birth spins of NS -- and thus of low-mass BHs -- to be small.  On the
other hand, observations of X-ray binaries (QPOs and iron lines)
suggests that higher-mass BHs ($M>5 M_\odot$) could be born with
moderate spin.
To allow for this
substantial uncertainty, in what follows we will consider the
implications of two models for the birth BH spin: $a=0$ and $a=0.5$.

\section{Mass accretion and spinup}
\label{sec:spinup}
Figure~\ref{fig:dM} demonstrates that the black hole accretes a
significant fraction of its mass during the HCE phase.
Since that matter probably carries some amount of angular
momentum, the black hole  could conceivably spin up significantly during
the HCE phase.
Unfortunately, the details of HCE accretion -- particularly at the
fine level of detail needed to resolve the amount of angular
momentum advected onto the compact object -- are not understood.

Lacking a quantitatively sound  choice for the specific angular
momentum, we argue instead that the specific  angular momentum accreted  should be {\em at most} 
the angular momentum of the marginally stable equatorial particle 
orbit.   This model, equivalent to assuming that the accretion
proceeds through a thin disk, implies that the final black hole spin
depends only on the total amount of matter accreted and the initial
black hole spin $J$ (or, as commonly denoted, $a=J/M^2$); as derived
initially by \citet{ThinDiskSpinup} \citep[see also][]{SpinPhotonLimit}, for a black hole which is initially 
nonspinning with mass $M_i$, the BH spin parameter $a$ is given by the
following expression $a=a_B(M,M_i)$:
\begin{equation}
\label{eq:bardeen}
a_B (M,M_i)=\left(\frac{2}{3}\right)^{1/2} \frac{M_i}{M} 
   \left[ 4- \left( \frac{18 M_i^2}{M^2} -2\right)^{1/2} \right].
\end{equation}
We consider this estimate to represent \emph{an upper limit} to the BH $a$,
since we do not expect that accretion in a HCE event will occur
through a thin disk.  Instead, we expect that on almost all scales the
accretion flow will be nearly radial.

\subsection{Accretion onto nonspinning holes}
We use our results on compact object masses at the onset and the end
of HCE to estimate $a$ under the thin disk assumption. We consider two
cases: (i) all mass accreted during the CE phases contributes to spinup and (ii) if accretion-induced collapse (AIC) occurs, only the fraction of the mass accreted after AIC contributes to spinup.

\begin{figure}
\includegraphics{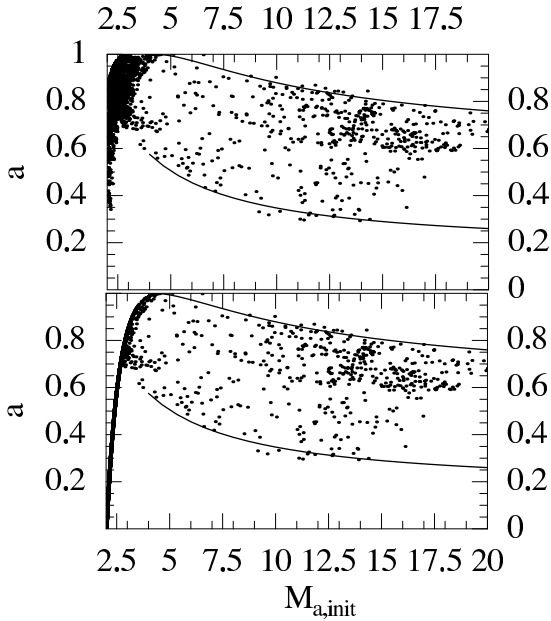}
\caption{\label{fig:spinup} For systems that undergo HCE evolution
  (i.e., the ``standard'' or ``reversed'' channels presented in
  Table~\ref{tab:main}, this plot shows the BH spin expected
assuming the HCE accretion occurs through a thin equatorial corotating
disk: points are
$(M_{a,\text{final}}, a(\Delta M))$ where $a$ is computed using the 
Bardeen formula [Eq.~(\ref{eq:bardeen})] assuming an initially
nonrotating hole.  The two panels show  the results assuming 
(i) all
the mass that is  accreted contributes to spinup 
and (ii) only the mass accreted after AIC of the
NS into a BH.  The solid curves (shown only for $M>4$) show the
empirical upper bound presented in Eq.~(\ref{eq:spinlimit}) for
nonspinning newborn BH.
}
\end{figure}

\emph{All mass contributes to spinup}:  As demonstrated in the top panel in 
Fig.~\ref{fig:spinup}, because of the substantial amount of mass
accreted during HCE phases, nearly all BHs could conceivably spin up
to very large $a$; a histogram of the values of $a$ we
observe would be highly concentrated near $a\sim 1$.  In particular,  many low-mass BHs (i.e, those
which formed via AIC during an HCE phase) have accreted of order half
their final mass, and can potentially spin up to of order $a\sim 1$.
Further, \emph{every} system that undergoes HCE evolution  accretes
enough mass to spin up 
significantly ($a>0.2$).
In this case, the BH-NS binaries with the largest spins (i.e., which
have $a\simeq 1$) have mass ratios
between $2:1$ to $3.5:1$.

However, for NS which undergo AIC, this approach approach will strongly overestimate the maximum
attainable spin, since it assumes disk accretion at the schwarzchild
radius, even before the BH forms.

\emph{Only post-AIC mass contributes to spinup}: Since current astrophysical evidence
suggests that NS do not get strongly spun-up during CE events, we
also consider the possibility that only the mass accreted {\em after}
the BH is formed 
through AIC contributes to its spin-up.   (Those BH which form through
direct collapse are treated as in the previous case.)  In this case, summarized in
the bottom panel in Fig.~\ref{fig:spinup}, we find a much broader
distribution of possible $a$ values (i.e., a histogram of possible $a$
outcomes would be more nearly flat, from $a=0$ to $a=1$). 
In this case, the systems with the largest spins have mass ratio near $3.5:1$.

Both models have the same behavior for systems with large final BH
masses  ($M_{\text{bh,f}}>4 M_\odot$).  Good empirical lower and upper bounds for $a$ for 
$M_{\text{bh,f}}>4 M_\odot$ follows
directly from the mass-accretion bounds 
Eqs.~(\ref{eq:HorribleGuess},\ref{eq:HorribleGuess2}):
\begin{eqnarray}
\label{eq:spinlimit}
a &<&  a_B \left(M_{bh,f},\frac{4 M_{bh,f}}{5} -8 M_\odot/5\right) \; \\
a &>&  a_B \left(M_{bh,f}, 0.95 M_{bh,f} - 0.57 M_\odot \right) \; 
\end{eqnarray}
where $M_{bh,f}$ and $M_{bh,i}$ are the BH final and birth masses,
respectively.

\subsection{Accretion onto spinning holes}
Black holes which are initially spinning present a more complex
accretion challenge in principle: the BH spin need not be
aligned with the disk angular momentum axis, so the accreted material could just as well
\emph{spin down} as spin up the hole.  However, following the same
process we used to circumvent the 
the (substantial) uncertainties we addressed earlier, we limit attention
to \emph{bounding the BH spin}.  The most conservative bound
is obtained by assuming a corotating equatorial disk.

\begin{figure}
\includegraphics{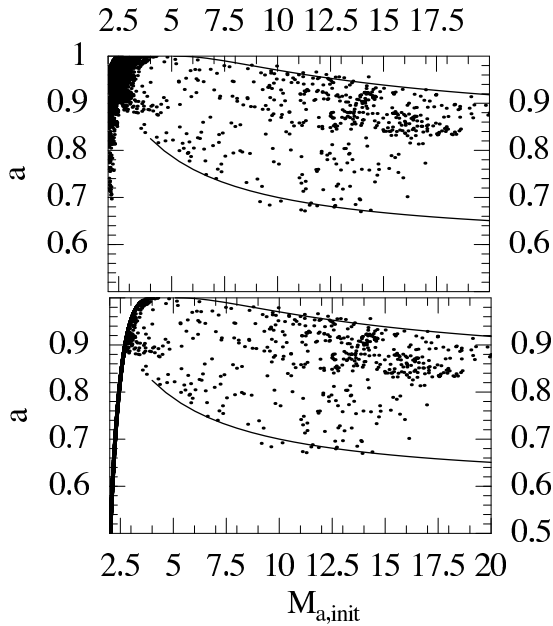}
\caption{\label{fig:spinupWithInitialSpin}
 As Fig.~\ref{fig:spinup}, 
but on a different vertical scale and assuming an initially
\emph{rotating}  hole: $a=0.5$. 
The solid curves (shown only for $M>4$) show the
empirical upper bound presented in Eq.~(\ref{eq:spinlimit2}).
 }
\end{figure}

Bardeen's formula [Eq.~(\ref{eq:bardeen})] applies equally well to BHs
which are initially spinning, if $M_i$ is chosen so the true initial
black hole mass $M_{bh,i}$ and spin $a_i$ satisfy $a_B(M_{bh,i})=a_i$.
To be concrete, if the birth BH spin is $a_i=0.5$, then choosing
\begin{equation}
M_i = 0.84 \;  M_{bh,i}
\end{equation}
yields $a_B(M_{bh,i}, M_i) =0.5$.  
When we repeat the analysis of the previous section, we find results
summarized in Figure~\ref{fig:spinupWithInitialSpin}.  
Since Bardeen's spinup relation is
monotonic, the empirical upper limit presented in
Eq.~(\ref{eq:spinlimit}) [for $a_i=0$] translates directly into a
corresponding spin limit when $a_i=0.5$
\begin{eqnarray}
\label{eq:spinlimit2}
 a &<&  a_B \left(M_{bh,f},0.84 \times\left[
   \frac{4 M_{bh,f}}{5} -8M_\odot/5\right]\right) \; 
   \; . \\
a &>&  a_B \left(M_{bh,f}, 0.84\times\left[
  0.95 M_{bh,f} - 0.57 M_\odot \right]\right)  \; 
\end{eqnarray}
Briefly, the
results are qualitatively similar to the nonspinning case, but
compressed in scale such that 
the minimum possible spin is $a=0.5$ instead of $a=0$.
Since the construction of spin as a function of BH mass is similar, the
spin upper bound also follows Eq.~(\ref{eq:spinlimit}), 
with the Bardeen formula calculated
choosing $M_i$ so the initial black hole has spin $a=0.5$.

\subsection{General Conclusions}
Our simulations have shown that almost all merging BH-NS binaries form
through a hypercritical common envelope (HCE) phase, where they accrete a
substantial fraction of their final mass.  Lacking a quantitative
model for the specific angular momentum accreted during HCE, 
we use the most pessimistic model: we assume the BH spins up through a
thin equitorial disk, according to the Bardeen formula
[Eq.~(\ref{eq:bardeen})].    Even setting aside those systems with
even greater physical uncertainties -- those systems with final BH
mass $M<4 M_\odot$, which form through accretion-induced collapse of
the NS and whose final spin depends additionally on the details of the
collapse and angular momentum transport onto the NS during that
collapse -- and even ignoring the possibility of BH birth spin,
we find that large spins ($a\sim 1$) could easily be
attained in principle from the HCE phase alone: with such a substantial
fraction of its final mass transferred, the BH could easily spin up.

However, for BHs with $M>4 M_\odot$, significant spinup after birth becomes
increasingly difficult, as larger black holes require ever larger (and
more unlikely) mass transfers in order to spin up to large $a$.
For example, assuming nonrotating initial BHs, the only BHs we found
with spin $a>0.8$ are those with final masses between $3 M_\odot$ and
$10 M_\odot$;
similarly, for BHs with $a=0.5$ at birth, the only BHs we found with 
spin $a>0.95$ are those with final masses in this same range.

\section{BH Spin and Detection of BH-NS Inspiral} 
In BH-NS binaries, the spin of the BH can, if large, have a
significant effect on the late stages of inspiral and in particular
the emitted gravitational waveforms
\citep[see][and references therein]{ACST,Apostolatos,Spiky3,BCVspin2,BCVspin3}.
As a result, in BH-NS binaries measurements of gravitational waves
offer the potential to extract information about the spin.
Furthermore, because of precession, such binaries will be much more
reliably detectable, suffering less frequently from the effects of
poor source orientation.

But these advantages come at a price.  The more complex waveforms of
spinning binaries are vastly more complicated to model.  At present
only ad-hoc models are considered computationally feasible
\citep{BCVspin2}, and even these methods involve substantial
computational challenges, particularly at more extreme mass ratios.

Our calculations provide extremely conservative upper bounds on the BH
spin in BH-NS binaries.  These bounds tell us that employing spin need
not pose such a dramatic dilemma, on astrophysical grounds: if we
employ spin, we could require fewer computational resources than we
might expect; and if we neglect it, we will not lose a dramatic
fraction of potentially observable events.

\subsection{Mass ratio spin constraints and detection}
When constructing template banks for BH-NS inspiral a priori, parameter ranges are
typically chosen to cover a very conservative parameter range; for
example, in one of the few papers that presented results for variable
mass ratio, \citet{BCVspin2} built a template bank based on 
all binaries with masses in the range
$M_{bh}\in[7,12]M_\odot$ and $M_{ns}\in[1,3]M_\odot$ and BH spins in
the range $a\in[0,1]$.  

However, most of the templates in the template bank
come from extreme-mass-ratio systems, involving a high-mass BH which
we have just demonstrated is difficult to spin up.  Therefore, a
different template bank, which neglected astrophysically irrelevant
systems, could perform the same search much more rapidly.
Detailed
computations of the computational savings we could obtain by
moderating our template bank are far beyond the scope of this paper,
and
will depend  sensitively on two poorly constrained
parameters, the BH birth spin and the maximum NS mass (here, assumed
$M=2 M_\odot$.)
However, in an optimistic case we expect that a more judicious choice of templates could
reduce the template bank size by a significant factor, 
cf. Sec. VI C of Pan et al.

\subsection{Limits on the decrease in event rate due to omitting BH
spin}
For the purposes of detection, spinning templates are required only
when nonspinning templates fail to have sufficient overlap with the
physical signal to
guarantee most inspiral events are detected.  
Loosely speaking, a nonspinning template fails to mimic the effect of
\emph{precession} on the waveform.  Because of spin-orbit coupling,
angular momentum will be exchanged, causing the orbital plane to
precess around the total angular momentum.  Therefore, spinning
templates are needed to detect those systems  where the effects
of precession are strongest -- namely, those with (i) strong BH spins
$a\approx 1$ that are (ii) strongly misaligned with the orbital
angular momentum \citep[see, e.g., Fig. 2 in][]{Spiky3}.

\citet{Spiky3} have quantitatively examined the expected overlap between nonspinning templates and
the expected signal from binaries in which the BH spin magnitude and
orientation are independently varied (see their Figure 2), assuming
BH-NS binaries with $M_{bh}=10 M_\odot$ and $M_{ns}=1.4 M_\odot$.  
They then
further convolved this function with an expected spin orientation
distribution to derive an estimate for the fraction of events that
could be seen that will be seen, as a function of spin magnitude of
the BH (see their Figure
3).\footnote{Specifically, they plot the average of the cube of the
  fitting factor versus the BH spin $a$.}  They find that even if all
BH were spinning maximally, only about $30\%$ of potentially-detectable
BH-NS mergers will be missed. 

The Grandclement et al. results, however, are not directly applicable
in our circumstances.  Spin-dependent modulations depend significantly
on mass ratio, a parameter that is varied in our  BH-NS binaries but
fixed for those of  Grandclement et al.  Nonetheless, if we employ
their relation between (i) the probability of detecting a detectable
source when ignoring spin and (ii) the BH spin magnitude of that
source, then we find when we convolve with our Monte Carlo sample of
 BH spins -- which, recall, was designed to produce the largest
 plausible spin values, given known mass transfer -- that we would lose at most
$30\%$ of events in extreme cases (i.e., high BH birth spin
or all HCE mass accreted onto NS) and slightly less ($25\%$) in more
conservative cases (i.e., mass before AIC does not contribute to spin).
Briefly, searches which use  nonspinning template banks will
likely find most detectable inspiral signals.  Practically speaking, 
given the considerable uncertainty in the underlying event rate
itself, upper limits produced using precise templates will have only
marginally more astrophysical impact than upper limits produced using
nonspinning template banks.
Of course, for \emph{parameter estimation}, correct templates are
essential in order to extract the BH spin, and thus to constrain its
accretion history.

\section{Summary and Future directions}
In this paper we perform population synthesis calculations to track the  history of systems which evolve into merging BH-NS
binaries, demonstrating that these objects form through only 
a few channels, almost exclusively involving
a hypercritical common envelope (HCE) phase and often involving
accretion induced collapse of a NS into a BH.  We show that a
significant amount of mass is typically accreted during this HCE
phase. 
Using \emph{ an upper limit based on thin-disk accretion}, we find the BHs
accrete enough matter to potentially spin up to  
large $a$ values, independent of the BH birth spin. 
Finally, we note  the observed sample of
neutron stars suggests the birth spins of the black holes in these
binaries should be small; therefore, the BH spins in BH-NS binaries
should arise almost entirely from the matter they accrete in a HCE phase.
Unfortunately, the HCE phase is poorly understood.
While our result is a first step
of a program to better understand the expected BH spins for those
systems
LIGO and other gravitational-wave detectors could observe, more
simulations of these processes 
are needed to obtain firmer conclusions.

Further, based on our rough understanding of the range of expected spin
magnitude, and using the
expected effect of spin on GW detection drawn from \citet{Spiky3}, we
demonstrated that neglecting spin in GW searches should only
moderately reduce the detection rate.  However, our analysis was based
on the work of \citet{Spiky3}, which like many papers in the
gravitational-wave literature assumed all BHs in BH-NS binaries had
$M=10 M_\odot$.  In fact, based on our calculations we expect (i) the
mass ratio can vary, and should be biased towards larger values (i.e.,
closer to $3:1$ than $10:1.4$); and (ii) the sample of BH-NS binaries
should show strong correlations between mass ratio and spin magnitude, because
higher-mass BHs are harder to spin up.  We hope to undertake a more
thorough Monte Carlo study of the effect of various expected distributions for BH
spin magnitude, spin tilt, and BH-NS mass ratio in a future paper.

Finally we note that in our simulations we consider only BH-NS in the
Galactic field;
we do not account for any stellar interactions relevant to
centers of globular clusters and similar dense environments. 
If BH-NS binaries form in
significant numbers in cluster centers, then their 
spin properties could be entirely unrelated to the analysis presented
in this study.

\acknowledgments

We thank Ron Taam, Fred Rasio, and Stu Shapiro for helpful
discussions. We particularly thank our anonymous referee for extensive comments and suggestions regarding the XRB literature.  
This work is partially supported by NSF Gravitational
Physics grants PHYS-0121416 and PHYS-0353111, a David and Lucile Packard Foundation Fellowship in Science and Engineering, and a Cottrell Scholar Award from the Research Corporation to VK. VK also thanks the Aspen Center for Physics for its hospitality while working on this paper.


\begin{thebibliography}{}
\bibitem[Apostolatos et al.(1994)]{ACST}  Apostolatos,  T. A.,   Cutler, C. , Sussman,  G. J., and  Thorne, K. S.\ 1994, Phys. Rev. D, 49, 6274
\bibitem[Apostolatos(1996)]{Apostolatos} Apostolatos, T. A.\  1996, Phys. Rev. D, 54, 2438
\bibitem[Bardeen(1970)]{ThinDiskSpinup} Bardeen, J. M.\ 1970,  Nature, 226, 64.
\bibitem[Beckwith and Done(2004)]{Beckwith} Beckwith, K., and Done, C.\ 2004, MNRAS, 352, 353.
\bibitem[Belczynski et al.(2002)]{StarTrack} Belczynski, K.,  Kalogera, V., and Bulik, T.   \href{http://adsabs.harvard.edu/cgi-bin/nph-bib_query?bibcode=2002ApJ...572..407B&amp;db_key=AST&amp}{2002, ApJ, 572, 407}
\bibitem[Belczynski et al.(2005)]{StarTrackUpdates} Belczynski, K., Kalogera, V., Taam, R., Rasio, F., Zezas, A.,  Bulik, T., Maccarone, T., Ivanova, N.\ 2005, ApJ, to be submitted
\bibitem[Brown et al.(2000)]{BrownHCE2} Brown, G. E., Lee, C.-H., and  Bethe, H. A.\ 2000,   ApJ, 541, 918.
\bibitem[Brown(1995)]{BrownHCE1} Brown, G. E.\  1995, ApJ, 440, 270.
\bibitem[Buonanno et al.(2004)]{BCVspin3}  Buonanno, A.,  Chen, Y.,  Pan, Y, and Vallisneri, M.\  2004, Phys. Rev. D, 70, 104003
\bibitem[Burgay et al.(2004)]{0737}    Burgay, M. et al.\ 2003, Nature, 426, 531
\bibitem[Duez et al.(2004)]{Duez1} Duez, M, Liu, Y. T., Shapiro, S, and Stephens,    \href{http://xxx.lanl.gov/abs/gr-qc/0402502}{gr-qc/0402502}
\bibitem[Kramer(2003)]{Kramer1} Kramer, M., et al.\ 2003, MNRAS, 342, 1299.  
\bibitem[Fryer et al.(1996)]{FryerBenzHerant} Fryer, C. L., Benz, W., and Herant, M.\ 1996, ApJ, 460, 801.
\bibitem[Grandclement et al.(2003)]{Mimic}  Grandclement, P.,  Kalogera, V.,  Vecchio, A.\ 2003, Phys. Rev. D, 67, 042003
\bibitem[Grandclement et al.(2004)]{Spiky3} Grandclement, P., Ihm, I,  Kalogera, V., and Belczynski, K.\ 2004, Phys. Rev. D, 69, 102002 
\bibitem[Hawke et al.(2004)]{Hawke1} Hawke, I., et al.\ 2004, Phys. Rev. D, 71, 024035  \href{http://xxx.lanl.gov/abs/gr-qc/0403029}{gr-qc/0403029}
\bibitem[Lindblom and Owen(2002)]{rmode1} Lindblom, L., and Owen, B.\ 2002, Phys. Rev. D, 65, 063006
\bibitem[Lorimer(2004)]{Lorimer1} Lorimer, D.\ 2004, Aspen Winter Conference on Astrophysics  (proceedings in press; available from    \texttt{www.astro.northwestern.edu/AspenW04/program.html})
\bibitem[Remillard et al.(2002)]{Remillard} Remillard, R., et al.\ 2002, ApJ, 580, 1030
\bibitem[Migliazzo et al.(2002)]{Mig1} Migliazzo et al.\ 2002, in  \emph{Neutron Stars in Supernova Remnants}, ASP Conference Series Vol. 271,   57
\bibitem[Miller et al.(2002)]{Miller1} Miller et al.\ 2002, ApJ, 570, L69
\bibitem[Miller et al.(2004a)]{Miller2} Miller et al.\ 2004, ApJ, 601, 45
\bibitem[Miller et al.(2004b)]{Miller3} Miller et al.\ 2004, ApJ, 601, L131
\bibitem[O'Shaughnessy et al.(2005a)]{RateFits}  O'Shaughnessy, R.,  Kalogera, V. , and Belczynsi, K.\ 2005,  ApJ, 620, 385
\bibitem[O'Shaughnessy et al.(2005b)]{PSconstraints}  O'Shaughnessy, R.,  Kim, C., Fragos, T., Kalogera, V., and Belczynski, K.\ 2005, ApJ, submitted
\bibitem[Pan et al.(2004)]{BCVspin2} Pan, Y,  Buonanno, A., Chen, Y.,  and  Vallisneri, M.\  2004,  Phys. Rev. D, 69, 104017
\bibitem[Portegies-Zwart and Yungelson(1998)]{PortZwart} Portegies-Zwart, S. F. \& Yungelson,  L. R.\ 1998,   A\& A, 332, 173
\bibitem[Rezolla et al.(2003)]{Rezolla} Rezolla, L.,  et al.\ 2003, MNRAS, 344, L34.
\bibitem[Shapiro(2002)]{ShapiroPolytropicCollapse}  Shapiro, S. and Shibata, M.\ 2002,  ApJ, 577, 904. \bibitem[Strohmayer(2001)]{Strohmayer} Strohmayer, T. E.\ 2001, ApJ, 552, L49
\bibitem[Thorne(1974)]{SpinPhotonLimit} Thorne, K. S.\ 1974, ApJ, 191, 507
\bibitem[Torok et al.(2005)]{Abramowicz} Torok, G., Abramowicz, M. A., Kluzniak, W., and Stuchlik, Z.\ 2005, A\&A 436, 1
\end{thebibliography}
\end{document}